\documentstyle[aps,epsfig,prl,multicol]{revtex}
\begin{document}
\title{Return probability: Exponential versus Gaussian decay}

\author{F.M.Izrailev and A.Casta\~neda-Mendoza}
\address{Instituto de F\'{\i}sica, B.U.A.P., Apdo. Postal J-48,
72570 Puebla, M\'exico}

\date{\today}
\maketitle

\begin{abstract}
We analyze, both analytically and numerically, the time-dependence
of the return probability in closed systems of interacting
particles. Main attention is paid to the interplay between two
regimes, one of which is characterized by the Gaussian decay of
the return probability, and another one is the well known regime
of the exponential decay. Our analytical estimates are confirmed
by the numerical data obtained for two models with random
interaction. In view of these results, we also briefly discuss the
dynamical model which was recently proposed for the implementation
of a quantum computation.
\end{abstract}

\begin{multicols}{2}


\section{Introduction}

As is known, in many-body systems the density of energy levels
increases extremely fast with an excitation energy. As a result,
for highly excited states even a very weak interaction between
particles can lead to a strong mixing between the unperturbed
basis states, resulting in a complex structure of exact
eigenstates. In this case one can speak about {\it chaotic
eigenstates} since their components can be practically treated as
pseudo-random ones. In the dynamics, this fact results in a
relaxation of the system to a steady-state distribution. The
studies of complex atoms \cite{FGGK94}, multi-charged ions
\cite{ions}, nuclei \cite{nuclei}, Bose-Einstein condensates
\cite{BBIS04}, and spin systems \cite{Nobel,spins} have confirmed
the dynamical origin of statistical laws in isolated systems (see,
also, Refs. \cite{FI97,I00} and the review \cite{kota}).

Recently, the theory of many-body chaos has been extended to the
models of a quantum computation. In particular, it was argued that
due to a very high density of energy levels, any kind of
perturbation may lead to decoherence effects thus destroying the
operability of a quantum computer. Therefore, it is of importance
to search for the conditions when the role of chaos can be
significantly reduced \cite{our,DISS05}.

So far, the study of the many-body chaos has been mainly
restricted by the investigation of statistical properties of the
energy spectra and eigenstates. On the other hand, in view of
experimental applications, one needs to know what are the {\it
dynamical} properties of quantum systems with strongly interacting
particles. Below, we analyze the time dependence of the widely
used quantity, the so-called return probability which determines
global properties of the dynamics. Our main interest is in the
interplay between the exponential and Gaussian decrease of the
return probability, in dependence on the strength of the
interaction between particles.

\section{Return probability and strength function}

In what follows we consider physical systems described by the
total Hamiltonian $H$ which can written in the separable form,
\begin{equation}
\label{H}H=H_0+V.
\end{equation}
Typically, such a representation is used when studying the
influence of a perturbation $V$ on a system governed by the
unperturbed Hamiltonian $H_0$. In many physical applications this
form of $H$ is quite natural, reflecting a different nature of
$H_0$ and $V$. However, quite often the separation of $H$ into the
two parts is not well defined, as in the case of mean field
approaches used to introduce ``good" variables in which $H_0$ has
a relatively simple form in comparison with a ``residual
interaction" $V$. In what follows we discuss some of generic
properties of the dynamics of the model (\ref{H}), by studying
specific forms of the Hamiltonian $H$.

It is naturally to represent $H$ in the unperturbed basis
$|k\rangle$ of $H_0$. Then the total Hamiltonian is presented by
the sum of the diagonal matrix $H_0$ plus the perturbation matrix
$V$ with the matrix elements $V_{lk}=\left<l|V|k\right>$. In order
to consider the evolution of wave packets in the basis of $H_0$,
one has to express {\it exact} eigenstates $\left| \alpha
\right\rangle $ of the total Hamiltonian $H$ in terms of {\it
basis} states $\left|k\right \rangle$ of $H_0$,
\begin{equation}
\label{exact} |\alpha \rangle = C_k^\alpha |k\rangle.
\end{equation}
The coefficients $C_k^\alpha$ give the expansion of an exact state
in terms of the basis states (for $\alpha$ fixed), or the
expansion of a basis state in terms of the exact states (for $k$
fixed). In principle, the knowledge of the state matrix
$C_n^\alpha$ and the corresponding energy spectrum $E^\alpha$
gives a complete information about the system.

Of particular interest is the case when the initial state
$\Psi(0)$ is a basis state $|k_0\rangle$. Then the evolution of
the $\Psi-$function is described by the expression,
\begin{equation}
\label{psit}\Psi (t) =\sum\limits_{n,k_0} C_n^\alpha
C_{k_0}^\alpha\left| k_0\right\rangle \exp(-iE^\alpha t).
\end{equation}
Here and below we assume that $\hbar=1$. As one can see, the
probability
\begin{equation}
\label{W0} w_k=|A_k|^2 =|\left\langle k|\Psi(t)\right\rangle|^2
\end{equation}
to find the system at time $t$ in the state $|k\rangle$ is
determined by the amplitude
\begin{equation}
\label{ampli} A_k= \left\langle k|\exp(-iHt)|k\right\rangle=
\sum\limits_\alpha|C_k^\alpha|^2\exp(-i E^\alpha t).
\end{equation}

Our main interest is in the {\it return probability} $W_0(t)$
which is the probability to find the system at time $t$ in the
initial state $|k_0\rangle$. One can see that the return
probability is determined by the expression,
\begin{equation}
\label{W00} W_0(t)\equiv w_{k_0}(t)=\left| A_{k_0}(t)\right| ^2,
\end{equation}
where
\begin{equation}
\label{A0} A_{k_0}(t)=\sum\limits_\alpha|C_{k_0}^\alpha|^2\exp(-i
E^\alpha t) \approx \int P_{k_0}(E)\exp(-i Et)dE.
\end{equation}
Here we replaced the summation by integration that can be done if
the number of large components $C_{k_0}^\alpha$ is large. Indeed,
these components strongly fluctuate around their mean values and
quite often can be considered as pseudo-random quantities. In
fact, this condition of a large number of pseudo-random components
in exact eigenstates can be used as the definition of chaos in
quantum systems (for details, see Refs.\cite{FIC96,FI97,I01}). In
this case, the time dependence of $W_0(t)$ is entirely determined
by the Fourier transform of $P_{k_0}(E)=P(E,E_{k_0})$ where $E$ is
the energy of exact eigenstates and $E_{k_0}$ is the energy
corresponding to the unperturbed state $|k_0\rangle$. This
quantity is known in the literature as the {\it strength function}
(SF) or {\it local spectral density of states},
\begin{equation}
\label{strength}P(E,E_{k_0})\equiv
\overline{|C_{k_0}^{(\alpha)}|^2}\rho (E).
\end{equation}
Here $\rho (E)$ is the density of states of the total Hamiltonian
$H$, and the average is performed over a number of states with
energies close to $E$.

To analyze generical properties of the return probability, let us
start with its behavior at small times. According to the
perturbation theory, one can easily get the general expression,
\begin{equation}
\label{smallT}W_0(t)\approx 1-\Delta_E^2t^2
\end{equation}
where $\Delta_E^2$ is the variance of the strength function in the
unperturbed energy space, determined as
\begin{equation}
\label{deltadef}\Delta_E^2 = \sum_{k\neq k_0} V_{k,k_0}^2.
\end{equation}
Note that the above expression is universal in the sense that it
is exact for any kind of the perturbation $V$. Practically, the
initial time scale for the perturbative expression (\ref{smallT})
to be valid is very small, and the main interest is in the
time-dependence of $W_0(t)$ beyond this time scale.

\section{Two-body random interaction model}

In order to analyze the behavior of $W_0(t)$ on a large time
scale, we consider the model which describes a closed system of
$N$ Fermi-particles occupying $M$ {\it single-particle} levels of
energies $\epsilon_s$. The total Hamiltonian can be represented in
the form (\ref{H}) where
\begin{equation}
\label{H0V} H_0 = \sum_{s=1}^{M} \epsilon_s\,a_s^{\dagger }a_s;
\:\:\:\ V = \sum_{s_1,s_2,s_3,s_4=1}^M
\tilde{V}_{s_1s_2s_3s_4}\,a_{s_1}^{\dagger }a_{s_2}^{\dagger
}a_{s_3}a_{s_4}.
\end{equation}
Here $H_0$ stands for non-interacting particles, and the
interaction $V$ between the particles is expressed in terms of
two-body matrix elements $\tilde{V}_{s_1s_2s_3s_4}$. The
many-particle basis $|k\rangle$ of $H_0$ is defined by the Slater
determinant, $\left| k\right\rangle =a_{s_1}^{\dagger
}\,.\,\,.\,\,.\,a_{s_{N}}^{\dagger }\left| 0\right\rangle$, where
$a_{s_j}^{\dagger }$ and  $ a_{s_j}$ are the creation-annihilation
operators, and $\left|0\right \rangle$ is the ground state. As one
can see, the interaction between particles is assumed to have a
two-body nature, therefore, each many-particle matrix element
$V_{lk}=\left<l|V|k\right>$ is a sum of a number of two-body
matrix elements $\tilde{V}_{s_1s_2s_3s_4}$ involving at most four
single-particle states $|s\rangle$ (for details, see, for
instance, Ref.\cite{I00}).

Note that the approach we consider here, is also valid for
quasi-particles that appear in the mean-field theories. In this
case $H_0$ stands for the mean-field part of the Hamiltonian, and
$V$ describes a residual interaction. The total Hamiltonian $H$ in
the form (\ref{H0V}) describes generic properties of such physical
systems as complex atoms, nuclei, quantum dots, etc. The energies
$\epsilon _s$ in such applications are, in fact, renormalized
quasi-particle energies.

In many realistic applications the interaction $V$ between
particles (quasi-particles) is so strong and complicated that
practically one can describe such an interaction by assuming that
all {\it two-body} matrix elements are distributed randomly
according to some distribution. Thus, the simplest version of the
two-body random interaction model (TBRI) is the Hamiltonian
(\ref{H0V}) in which all matrix elements
$\tilde{V}_{s_1s_2s_3s_4}$ are random Gaussian numbers with the
zero mean and the variance $\left\langle
\tilde{V}^{2}\right\rangle $. It is interesting to note that the
{\it many-body} matrix elements $V_{lk}$ are, however, weakly
correlated, due to the fact that the same two-body matrix elements
enter in different {\it many-body} matrix elements $V_{lk}$. In
general, these correlations can be neglected, however, for
specific observables they give rise to unexpected results (for
details see Ref.\cite{FGI96}).

Without the loss of generality one can assume that the
single-particle spectrum has the constant mean level spacing,
$d_0=\,\langle\epsilon _{s+1}-\epsilon _s\rangle=1$; here the
brackets $\langle . . . \rangle$ stand for the average over random
values of $\epsilon_s$. The number of many-body states increases
very fast with an increase of the number of particles $N$ and
number $M$ of single-particles states. For this reason even for a
relatively small number of particles the exact eigenstates may
consist of many unperturbed basis states, thus providing us with a
possibility to use statistical methods. In particular, a novel
approach has been developed in Refs. \cite{FI97,FIC96,FI99}, that
is based on the chaotic structure of eigenstates in a given basis
of unperturbed many-particle states. This approach allows one to
relate statistical properties of exact eigenstates in many-body
representation directly to the properties of single-particle
operators, such as the occupation number distribution of
single-particle states.

As was shown above, the time dependence of the return probability
is entirely determined by the Fourier transform of the SF. In many
applications the SF is known to have the Breit-Wigner (BW) form
(or, the same, the Lorentzian) resulting from the application of
the Random Matrix Theory, (see, for instance, Ref. \cite{BM74}).
The half-width $\Gamma_0$ of the BW is given by the Fermi golden
rule and in our case reads as
\begin{equation}
\label{Gamma0}\Gamma_0(E)\equiv \Gamma(E,k_0) \simeq 2\pi
\overline{\left| V_{k_0k}\right| ^2} \rho_0(E).
\end{equation}
Here $\overline{\left| V_{k_0k}\right| ^2}$ is the mean square
value of many-body matrix elements (obtained by the average over
$k$), and $\rho_0(E)$ is the density of those states which are
directly coupled to the basis state $|k_0\rangle$ by the
interaction $V$. Note that this density $\rho_0(E)$ is much
smaller than total density $\rho_{tot}(E)$ of all many-body
states. This fact is manifested by a large number of zero matrix
elements for any fixed line in the Hamiltonian matrix $H_{lk}$.

It should be stressed that the above result for the BW form of the
SF is based on a non-perturbative approach according to which a
large number of many-body states are coupled by a relatively
strong interaction. As a result, the decrease of the return
probability has the exponential time dependence,
\begin{equation}
\label{Lor} W_0(t)\simeq \exp\left(-\Gamma_0 t\right),
\end{equation}
apart from a small time scale $t < t_0$ on which the quadratic
decrease (\ref{smallT}) occurs.

For a long time it was assumed that the exponential decrease of
the return probability is the only regime which is physically
relevant to the dynamics of systems with many interacting
particles. However, recently it was found that in many situations
the form of the SF can be quite close to the Gaussian (see, e.g.
\cite{nuclei}). This fact is related to a finite width $\Delta_V$
of the interaction in the energy space for isolated systems. For
the TBRI model it was shown \cite{FI01b} that if $\Gamma_0$ is
much less than $\Delta_V$, the form of the SF, is, indeed, the BW.
However, in the other limit, $\Gamma_0 \geq \Delta_V$, of a very
strong interaction, the leading dependence of $W_0(t)$ is the
Gaussian,
\begin{equation}
\label{gau} W_0(t) \simeq \exp (-\Delta_E^2t^2),
\end{equation}
and occurs on a time scale $0 < t \leq t_c $. After, for $t
> t_c $, the decay of $W_0(t)$ is described by the exponential function
\cite{FI01a}.

The transition from the BW to the Gaussian for the TBRI model has
been analyzed in Ref.\cite{FI00}, although the analytical
expression in a closed form is unknown. In order to evaluate
$W_0(t)$, in Ref.\cite{FI01a} a phenomenological expression was
suggested that depends on both parameters, $\Gamma_0$ and
$\Delta_E$. The analytical expression for the variance
$\Delta_E^2$  of the strength function can be found explicitly
\cite{FI97},
\begin{equation}
\label{Delta_E} \Delta_E^2=\frac{v_0^2 }{12}N(N-1)(M-N)(M-N+3).
\end{equation}
Here $[-v_0, v_0]$ is the range within which the two-body matrix
elements are distributed randomly with a constant probability,
therefore, $\left < \tilde{V}^2\right >=v_0^2/3$. It is
interesting to note that for Fermi-particles the variance
$\Delta_E^2$ turns out to be independent of a specific basis state
$\left|k_0 \right \rangle$.

Thus, in the case of a not very strong perturbation the decrease
of the return probability is the exponential one, and with an
increase of the interaction $V$, one should expect a quite large
time scale on which the Gaussian form (\ref{gau}) occurs. Let us
check these predictions by making use of numerical data. In all
our calculations we have used $N=6$ Fermi-particles occupying
$M=12$ single-particle states; this results in the size $924
\times 924$ of the Hamiltonian matrix. For simplicity, the
initially excited state was taken at the center of the energy
spectrum, $k_0=462$, where the density of many-body states is
maximal, and the energy spectrum is symmetrical.

For a relatively weak (however, non-perturbative) interaction,
$v_0=0.12$, the data, indeed, demonstrate a clear exponential
dependence, up to some time scale beyond which the finite size
effects of the Hamiltonian matrix are important, see Fig.1. The
fit to the exponential dependence (\ref{Lor}) gives $\Gamma_0
\approx 0.97$. To compare with the analytical expression
(\ref{Gamma0}), one should note that for the TBRI model this
expression is difficult to use directly, since the quantity
$\Gamma_0(E)$ is not well-defined by Eq.(\ref{Gamma0}). The
problem is that the density of directly coupled many-body states
strongly changes in dependence on $k$. The rough estimate can be
obtained as follows. First, one can relate the term
$\overline{\left| V_{k_0k}\right|}^2$ to $\Delta_E^2$ in the way,
$\Delta_E^2=K \overline{\left| V_{k_0k}\right|} ^2$, where $K$ is
the number of non-zero elements in any line of the Hamiltonian
matrix (which is independent of $k_0$ \cite{FI97}). Second, the
simplest estimate of the mean density $\rho_0$ is due to the
average width $\Delta_V$ of the interaction in the energy space,
$\rho_0 \approx K/\Delta_V$. As a result, one gets,
\begin{equation}
\label{Gamma_est} \Gamma_0 \approx 2\pi
\frac{\Delta_E^2}{\Delta_V}.
\end{equation}
Finally, it can be shown that the simplest estimate for the width
of the interaction reads as $\Delta_V \approx 2 d_0 (M-N)$. This
gives $\Gamma_0 \approx 1.03$ which is a good result, taking into
account the problems with the evaluation of the expression for
$\Gamma_0$.

\begin{figure}[htb]
\begin{center}
\hspace{-1.0cm} \epsfig{file=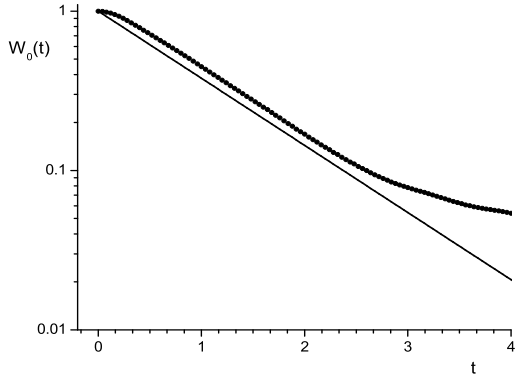,width=3.0in,height=2.3in}
\caption{Time dependence of the return probability $W_0(t)$ for
the interaction strength $v_0=0.12$ for which the SF has the BW
form. Full circles stand for numerical data, and the straight line
(shifted for a better visualization) is the linear fit on the time
scale where the linear slope is clearly seen.}
\end{center}
\end{figure}

\begin{figure}[htb]
\vspace{-1.5cm}
\begin{center}
\hspace{-1.0cm} \epsfig{file=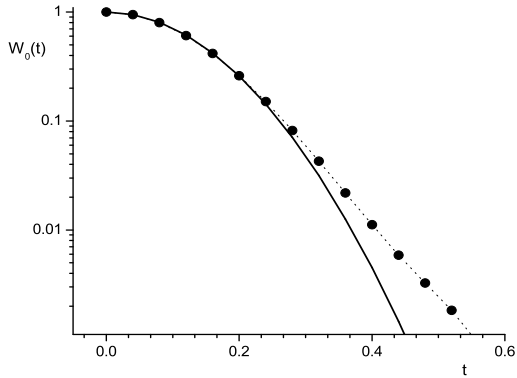,width=3.0in,height=2.3in}
\caption{ Time dependence of the return probability $W_0(t)$ for
the case of the Gaussian form of the strength function. Full
circles correspond to the numerical data for $v_0=0.5$ and the
solid curve is the analytical dependence (\ref{gau}) with
$\Delta_E^2$ determined by Eq.(\ref{Delta_E}).}
\end{center}
\end{figure}

Now let us consider another limit case of a very strong
interaction, $v_0=0.5$, when the SF is quite close to the
Gaussian. Numerical data reported in Fig.2 manifest a long
Gaussian decrease of the return probability. Note that the
deviation from the Gaussian dependence towards the exponential one
(linear slope in Fig.2 after $t\approx 0.4$) starts for very small
values of $W_0(t)$. Therefore, practically the decrease of the
return probability is described by the dependence (\ref{gau}).

Finally, we conclude with the intermediate case when both
dependencies, (\ref{gau}) and (\ref{Lor}), are important for the
description of the return probability. In Fig.3 one can see that
there are two time scales. On the first time scale, $t \leq t_c
\approx 0.3$,  the decrease of $W_0(t)$ has the Gaussian form
(\ref{gau}), and for $t > t_c$ it changes to the exponential
dependence. It is now instructive to analyze the critical time
$t_c$ which divides these two regimes,
\begin{equation}
\label{critical} t_c \approx \frac{\Gamma}{\Delta^2_E} \approx
\frac {2\pi}{\Delta_V}.
\end{equation}
One can see that $t_c$ is the time to resolve the finiteness of
the width of the interaction. If this width is very large, the
exponential decrease starts on a small time scale. Contrary, for
relatively small values of $\Delta_V$ the Gaussian decrease of
$W_0(t)$ starts from $t=0$ and lasts for a long time. According to
this estimate, we have $t_c \approx 0.5$ which roughly corresponds
to the data. Note that the critical time $t_c$ is, in fact, not a
well defined quantity and can be determined up to some numerical
factor of the order one. Remarkably, the time of the
correspondence of the data to Eq.(\ref{gau}) turns out to be
independent of the perturbation strength.

\begin{figure}[htb]
\begin{center}
\hspace{-1.0cm} \epsfig{file=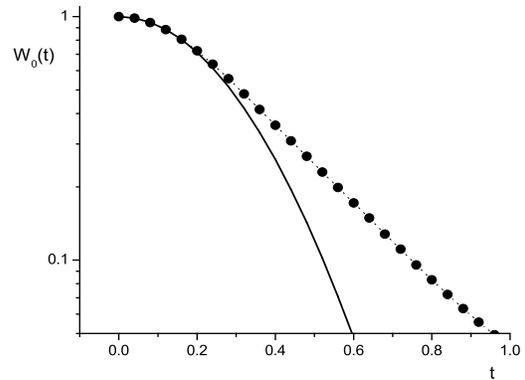,width=3.0in,height=2.3in}
\caption{Return probability $W_0(t)$ for the intermediate
situation with $v_0=0.25$, when both regimes, the Gaussian and
exponential ones, are characteristic of the dynamics. Full circles
stand for the numerical data, and solid curve is the theoretical
expression (\ref{gau}).}
\end{center}
\end{figure}

\section{Wigner band random matrices}

It is instructive to apply the above analysis to the so-called
Wigner band random matrices (WBRM). These matrices are very useful
for understanding generic properties of realistic physical systems
of interacting particles. The WBRM model is described by the
Hamiltonian (\ref{H}) which consists of two parts, one of which is
a diagonal matrix $H_0$ with increasing entries $\epsilon_j$. This
part can be treated as the ``mean field" part of the total
Hamiltonian $H$. Another part is a banded matrix $V_{ij}$, which
is associated with the interaction between unperturbed basis
states. Thus, the model has the following form,
\begin{equation}
\label{wbrm}H_{ij}=\epsilon _j\delta _{ij}+V_{ij},
\end{equation}
where $\delta_{ij}$ is the delta-function. It is assumed that
random values $\epsilon _j$ with the mean spacing $D$ are
reordered in an increasing way, $\epsilon_{j+1} > \epsilon_j$. As
for the off-diagonal matrix elements $V_{ij}$, they are
distributed according to the gaussian distribution (with the zero
mean, $ <V_{ij}>=0$, and the variance $<V_{ij}^2>=V_0^2$) for the
matrix elements inside the finite band $|i-j|\le b/2$, and zero
otherwise.

These matrices have been introduced by Wigner in Ref.\cite{W55} in
application to nuclear physics. A particular interest was the form
of the strength function in dependence on the strength of the
interaction $V$. It was shown that the form of the SF is the BW
for a moderate (non-perturbative) strength $V$, and the semicircle
for a very strong perturbation. Full analytical treatment of the
form of the SF for Eq.(\ref{wbrm}) is given in Ref.\cite{FCIC96}
with the use of the modern approach. In particular, it was found
that in the transition from the BW to the semicircle, the form of
the SF is very close to the Gaussian.

The condition for the SF to be of the Breit-Wigner form in the
WBRM model can be written as follows \cite{FCIC96},
\begin{equation}
\label{range1}D \ll \,\Gamma_0 < \Delta_V;\,\,\,\,\,\, \Delta_V=b
D=b\rho_0^{-1},
\end{equation}
where the half-width $ \Gamma_0$ is given by the Fermi golden
rule,
\begin{equation}
\label{BWgam}\Gamma_0=2\pi \rho _0V_0^2,
\end{equation}
and $\Delta_V$ is the energy width of the interaction $V$.

The left part of the inequalities in Eq.(\ref{range1}) indicates
the non-perturbative situation for which many of unperturbed basis
states are strongly coupled by the interaction. On the other hand,
the interaction should not be very strong, namely, the width
$\Gamma_0$ determined by Eq.( \ref{BWgam}), has to be less than
the width $\Delta_V$ of the interaction in the energy
representation. The latter condition is generic for systems with
finite range of the interaction $V$. One should stress that,
strictly speaking, the BW form of the strength function is not
correct in physical applications since its second moment diverges
(which assumes an infinite range of the interaction). For the WBRM
model with finite values of $b$ it was shown \cite{W55,FGGK94}
that far off the energy range $ \Delta_V$ the SF decreases faster
than a pure exponent.

In contrast with the TBRI model, in the WBRM model the energy
scale $\Delta_V$ is well defined that simplifies our further
analysis. One can see that instead of the control parameter
$\Delta_V$, one can equivalently use the variance $\Delta _E^2$ of
the SF, which can be expressed through the off-diagonal matrix
elements of the interaction, $ \Delta _E^2=\sum_jV_{ij}^2$ for
$i\neq j$, therefore, $\Delta _E^2=bV_0^2$. As a result, we have
$\Delta _V=2\pi \Delta _E^2/\Gamma_0$ and the relation
(\ref{range1}) can be written in the form,
\begin{equation}
\label{range2}D\ll \,\Gamma_0 < \Delta _E \sqrt{2\pi }.
\end{equation}

Numerical data for the WBRM model have confirmed that for
$\Gamma_0\approx 2\Delta _E$ the form of the SF is quite close to
the Gaussian. Moreover, it was found that the transition from the
BW dependence to the Gaussian-like is very sharp. Note that the
extreme limit of a very strong interaction, $\Gamma_0\gg 2\Delta
_E$, seems to be non-physical, giving rise to the semicircle form
of the SF. The same effect occurs for the TBRI model for which the
relation $V\gg H_0$ means that the residual interaction is much
stronger than the mean field part.

Applying the arguments given for the TBRI model, the Gaussian
decrease of the SF has the form,
\begin{equation}
\label{WW} \ln W_0(t) = - \Delta_E^2 t^2
=-\Gamma_{0}\frac{t^2}{t_c} = - \frac{4\pi^2 V_0^2}{bD^2} \xi^2
\end{equation}
with $\xi=t/t_c$ and $t_c=2\pi/bD$.

\begin{figure}[htb]
\begin{center}
\hspace{-1.0cm} \epsfig{file=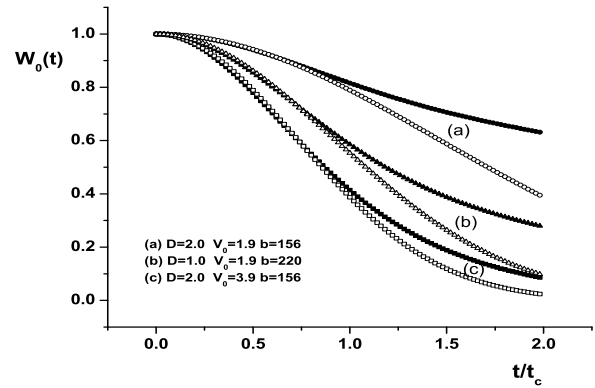,width=3.4in,height=2.3in}
\caption{Return probability $W_0(t)$ for the WBRM model
(\ref{wbrm}). Full and open symbols stand for numerical data and
theoretical expression (\ref{WW}), respectively.}
\end{center}
\end{figure}

Numerical data in Fig.3 obtained for $N=924$ and different model
parameters, are in a good correspondence with our analysis. First,
the time dependence of $W_0(t)$ on the time scale $t \lesssim t_c
$ is, indeed, of the Gaussian form (\ref{WW}). Second, the
critical time which divides the two characteristic dependencies,
nicely corresponds to the expression $t_c \approx 2\pi/bD \approx
1.0$. As one can see, the transition from the Gaussian to the
exponential decrease is a quite generic property of the wave
packet dynamics in the systems which are described by the TBRI or
WBRM models.

\section{Dynamical model of a quantum computation}

Now we apply our analysis to a physical model of quantum
computation which has no random parameters. Although a direct
application of the results obtained for random matrix models to
the dynamical models is not justified, it is of interest to see
whether such a comparison is possible. The model we consider here
was recently proposed \cite{ber1} as a simple realization of a
solid-state quantum computation. It describes a one-dimensional
chain of $L$ interacting $1/2$-spins (qubits) that are subject to
an external magnetic field. In order to have selective resonant
excitation, the time independent part $B^z = B^z(x)$ of a magnetic
field is assumed to have a constant gradient along the
$x$-direction. This provides different Larmor frequencies for
different spins, $\omega_k = \gamma_n B^z=\omega_0+ak$, where
$\gamma_n$ is the spin gyromagnetic ratio and $a$ is proportional
to the gradient of the constant part of the magnetic field (see
details in \cite{our}).

For a specific pulse of the time-dependent part of the magnetic
field, resulting in a single cubit operation, one can derive the
time-independent Hamiltonian (\ref{H}) with
\begin{equation}
\label{HV} H_0=\sum_{k=0}^{L-1} \Big [-\xi_k  I^z_k- 2J I^z_k
I^z_{k+1} \Big] \,;\,\,\,\,\,\,\,\,\,\,\,\,\, V= -
\sum_{k=0}^{L-1} \Omega I^x_k,
\end{equation}
where $\xi_k=a k$ \cite{our}. Here the frequency $\Omega$ is the
Rabi frequency of the $p$-th pulse, $I_k^{x,y,z} = (1/2)
\sigma_k^{x,y,z}$ with $ \sigma_k^{x,y,z}$ as the Pauli matrices,
and $I_k^{\pm}=I^x_k \pm iI^y_k$. It is also assumed that the
interaction $J$ between nearest qubits does not depend on the
indexes $k$ and $k+1$.

The unperturbed basis (in which $H_0$ is diagonal) is reordered
according to an increase of the index $s$ which is written in the
binary representation, $s=i_{L-1},i_{L-2},...,i_{0}$ (with $i_s=0$
or $1$, depending on whether the single-particle state of the
$i-$th qubit is the ground state or the excited one). Therefore,
the parameter $\Omega$ corresponds to a non-diagonal coupling,
thus, determining the matrix elements $V_{kn}=V_{nk}=- i\Omega/2$
with $n \neq k$. As one can see, in contrast with the TBRI model
discussed above, in the chosen representation the interaction
between particles is absorbed by $H_0$, and $V$ describes the
coupling to the external magnetic field.

The problem studied in Refs.\cite{our} was the analysis of whether
the inter-qubit interaction, as well as the interaction of qubits
with the external magnetic field, can be a source of a kind of
internal decoherence caused by quantum chaos. It is a wide-spread
concern that for many interacting qubits the onset of quantum
chaos may occur even for a very weak interaction, see, for
instance, \cite{spins,dima}. This expectation is based on the fact
that with an increase of number of qubits the level density of
many-body states increases drastically, thus strongly enhancing
the delocalization effects due to the interaction between qubits.
The simple estimate \cite{spins} shows that generically the
threshold for the onset of chaos decreases as $J_{cr} \sim 1/L$
where $J_{cr}$ is a critical inter-qubit interaction above which
the eigenstates are extended over unperturbed many-body states.

However, as was found in Refs.\cite{our}, in the case of an
external magnetic field with a constant gradient along the chain
of qubits, the onset of chaos is strongly suppressed. Although in
this case the total energy width may be large (proportional to
$L$), the estimate \cite{BBIT01a} shows a feasibility of an
experimental realization for a quite large number of qubits.
Another principally different scheme that allows to avoid strong
delocalization/chaos effects, is suggested and analyzed in
Refs.\cite{DISS05} for a quantum computer based on electrons on
helium.

Our interest below is to see whether the results for the return
probability, obtained above in terms of random matrix models, can
be applied to the dynamical model (\ref{HV}). Note that a strong
decrease of the return probability $W_0(t)$ can serve as an
effective measure of a stability of a quantum computation. It can
be shown that the delocalization effects are directly governed by
the time-dependence of $W_0(t)$. Strictly speaking, the
effectiveness of a quantum computation should be analyzed for the
time-dependent model with a large number of pulses. However, if
for a single pulse the effects of quantum chaos are strong, they
will be generically enhanced in the presence of many pulses. This
is why below we restrict ourselves by a consideration of the
simplified time-independent Hamiltonian (\ref{HV}). We hope that
our analysis may have also an interest in view of general problems
of the dynamics of systems with a large number of interacting
spins.

First, we start with the so-called {\it non-selective} regime
which is defined by the conditions, $\Omega\gg a \gg J$. This
inequality provides the simplest way to prepare a homogeneous
superposition of $2^L$ many-body states, which is needed to start
with the implementation of the Shor's or Grover's algorithm. The
analytical and numerical treatment of the model (\ref{HV}) in this
regime has revealed \cite{our} that the constant gradient magnetic
field (with $a\neq 0$) strongly reduces unwanted effects of
quantum chaos. Specifically, it was shown that in this case the
chaos border turns out to be independent of the number $L$ of
qubits, in contrast to the models thoroughly studied before
\cite{dima}. In particular, the quantum chaos may occur only for a
large coupling $\Omega$ and strong interaction $J$ between qubits.
Another new effect which was found in Refs.\cite{our}, is that the
border of quantum chaos does not coincide with the border of
delocalization. This peculiarity is important in view of
applications to integrable or nearly integrable models for which
the quantum chaos is absent, however, the delocalization effects
can be very strong.

As was shown in Refs.\cite{our} in the non-selective regime with
$\Omega=100$, the delocalization effects start to play essential
role for $J \geq J_c \approx 10$. On the other hand, the quantum
chaos effects, such as the Wigner-Dyson distribution for the
spectra statistics, occurs for $J \geq J_q \approx 100$. Since in
this case both the inter-qubit interaction and the perturbation
due to the magnetic field are strong, one can analyze the dynamics
of the return probability $W_0(t)$ in connection with the
previously discussed results.

\begin{figure}[htb]
\begin{center}
\hspace{-1.0cm} \epsfig{file=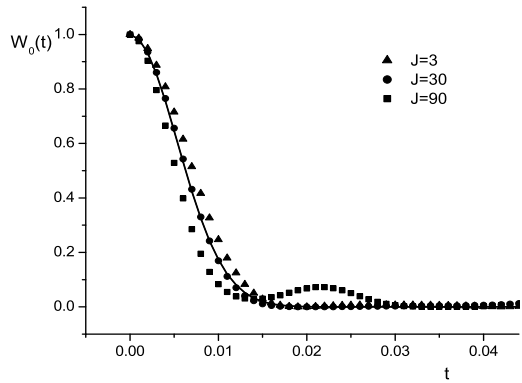,width=3.0in,height=2.3in}
\caption{Return probability $W_0(t)$ for the dynamical model
(\ref{HV}) of a quantum computation for $L=8, \Omega=100, a=1$ and
different values of $J$. Solid curve is the theoretical expression
(\ref{WW}) for the Gaussian dependence, and full squares,
triangles, circles are numerical data.}
\end{center}
\end{figure}

Numerical data presented in Fig.5 show a strong decrease of
$W_0(t)$ for three values of $J$. As one can see, in all cases the
return probability decreases similar to the Gaussian form of
Eq.(\ref{gau}). The correspondence with theoretical predictions
should be treated as a good one, taking into account the dynamical
character of the considered model. It is important that the
perturbation $\Omega$ due to the magnetic field is strong,
therefore, the results are almost insensitive to the inter-qubit
interaction in a very wide region of $J$.

Now we analyze the regime of {\it selective excitation} which is
characterized by the following range of parameters, $\Omega \ll J
\ll a $. In this regime each pulse acts selectively on a chosen
qubit thus resulting in a resonant transition. In
Ref.\cite{BBCo02} this regime was analyzed in detail with the main
interest to the fidelity of some quantum protocol (for the
time-dependent Hamiltonian with many pulses). It was shown that in
this regime for a relatively large gradient of the magnetic field
there is no any danger of quantum chaos, and the dynamics can be
analyzed on the base of the perturbation theory. It is instructive
now to see how the return probability $W_0(t)$ corresponds to
these results.

\begin{figure}[htb]
\begin{center}
\hspace{-1.0cm} \epsfig{file=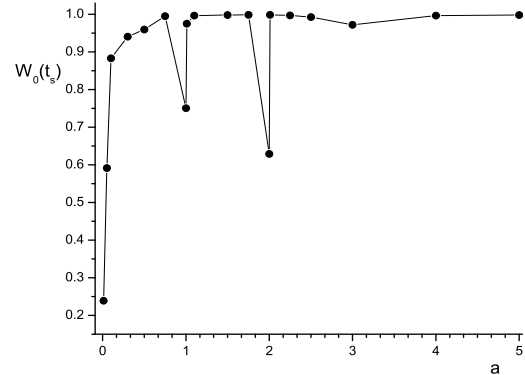,width=3.0in,height=2.3in}
\caption{Return probability $W_0$ at time $t_s \approx 3.0$ for
the model of quantum computation with $L=8, \Omega=0.118, J=1$, in
dependence on the parameter $a$ which is proportional to the
gradient of the magnetic field.}
\end{center}
\end{figure}

In Fig.6 we present numerical data demonstrating the dependence of
$W_0$ for some time $t_s \approx 3.0$. First, one can see that for
small values of $a$ the perturbation turns out to be very strong,
leading to a strong decrease of $W_0$. This result demonstrates
that even a small interaction between qubits gives rise to a
strong leakage of the probability from an initially excited state.
The origin of this phenomenon is the degeneracy (for $a=0$) or
quasi-degeneracy (for small $a$). With an increase of the gradient
of the magnetic field, the dynamics turns out to be very stable,
as is manifested by the values of $W_0(t_s)$ close to one. The
analytical estimate of the critical value of the inter-qubit
coupling above which the fidelity is very high, was found
\cite{BBCo02} to be $a_{cr} \approx 4 J $. Our data, however,
clearly show a slightly different value $a_{cr} \approx J$. A more
careful inspection of the numerical data of Ref.\cite{BBCo02}
confirms that, indeed, the fidelity starts to decrease at smaller
than $a \approx 4$ values. Our data in Fig.5 also reveal a quite
interesting resonance effect that occurs for specific values $a=1$
and $a=2$. This effect has the same origin as that found in
Ref.\cite{BBCo02}.

In order to see more clearly the role of the gradient magnetic
field, we performed an additional check of the time dependence of
the return probability for small, $a=0.05$,  and large, $a=3.0$,
values of $a$. The results in Fig.7 manifest a very different
behavior of the return probability for these two representative
values. If for the small value of $a$ the decrease of $W_0(t)$ is
very fast and seems to be non-recurrent, for large $a=3.0$ the
return probability remains very close to one and shows a clear
recurrence. It is interesting to note that for $a=0.05$ one can
see two regimes, the Gaussian one (for $t < 2.0)$, and the
exponential one (for $t > 3.0$).  Note that due to a specific
character of the selective excitation (quasi-degeneracy for
$a=0.05$ and very weak perturbation for $a=3.0$), the comparison
with the discussed above analytical estimates in this case is not
valid.

\begin{figure}[htb]
\begin{center}
\hspace{-1.0cm} \epsfig{file=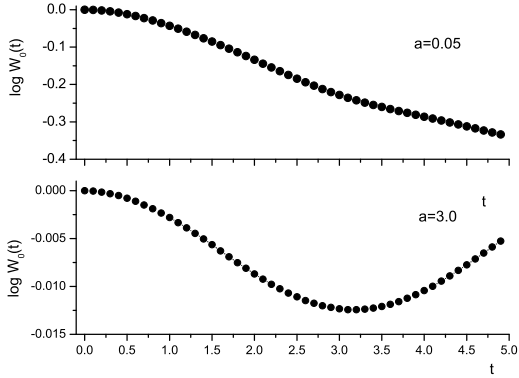,width=3.0in,height=2.3in}
\caption{Return probability $W_0(t)$ for small, $a=0.05$, and
large, $a=3.0$, values of the parameter $a$, for the parameters of
Fig.6.}
\end{center}
\end{figure}

\section{Conclusion}

In conclusion, we have studied the return probability $W_0(t)$ in
three models of strongly interacting particles. The first model is
the model with random two-body interaction, typically used to
describe the many-body systems of Fermi-particles such as heavy
nuclei, many-electron atoms, quantum dots, etc. As usual, the
two-body random matrix elements are assumed to be gaussian random
entries, which is known to be reasonable when the interaction is
strong and has a complicated form. The main result of our
analytical and numerical study is that with an increase of the
interaction, the Gaussian decrease of the return probability can
last for a long time. The typical picture is the following: on
some time scale $t < t_c$ the decrease of $W_0(t)$ is the
Gaussian, and for $t > t_c$ it is the exponential one. The
Gaussian decrease on the scale $0 < t < t_c$ can be either weak or
strong, depending whether the interaction is strong or very
strong. Thus, the standard exponential decrease, associated with
the Fermi golden rule, is not correct for a very strong
perturbation. Numerical data confirm the analytical predictions.

In order to elucidate the meaning of our analytical estimates, we
have also analyzed the model of Wigner band random matrices which
captures essential features of the chaotic systems of interacting
particles. This model turns out to be very effective since it
demonstrates in a very transparent way the dependence of the
dynamics on few global parameters of physical significance. Our
numerical data reflect the generic properties of the interplay
between the Gaussian and exponential decrease of the return
probability.

Finally, we have analyzed the model which was recently proposed as
an implementation of a quantum computer. In contrast with the
random matrix models, this model is purely dynamical one (without
any random parameters). As is now well understood, the effects of
quantum chaos and/or delocalization may also arise in the models
of quantum computation, and lead to a kind of internal
decoherence. Our numerical data show that in the region of
parameters that corresponds to a strong delocalization and quantum
chaos, the return probability has the Gaussian form of the decay
for a quite a long time, similar to what occurs in the random
matrix models. Additional study of the role of a non-zero gradient
magnetic field has shown that the fast decrease of the return
probability corresponds to earlier results for the fidelity of
some quantum protocol with many pulses. Therefore, the return
probability can be considered as a quite sensitive quantity for
establishing the region of parameters where one can expect stable
operability of quantum computers.

This research was partially supported by the CONACYT (M\'exico)
grant No~43730.

\end{multicols}

\begin{thebibliography}{99}

\bibitem{FGGK94}  V.V.~Flambaum, A.A.~Gribakina, G.F.~Gribakin, and M.G.
~Kozlov, Phys. Rev. A 50 (1994) 267.

\bibitem{ions}  G.F.~Gribakin, A.A.~Gribakina, V.V.~Flambaum, Aust.
J. Phys. 52 (1999) 443.

\bibitem{nuclei}  M.~Horoi, V.~Zelevinsky and B.A.~Brown, Phys. Rev. Lett.
74 (1995) 5194; V.~Zelevinsky, M.~Horoi and B.A.~Brown, Phys.
Lett. B 350 (1995) 141; N.~Frazier, B.A.~Brown, and V.~Zelevinsky,
Phys. Rev. C 54 (1996) 1665; V.~Zelevinsky, B.A.~Brown, M.~Horoi,
and N.~Frazier, Phys. Rep. 276 (1996) 85.

\bibitem{BBIS04} G.P.~Berman, F.~Borgonovi, F.M.~Izrailev, and A.~Smerzi,
Phys. Rev. Lett. 92 (2004) 030404.

\bibitem{Nobel}  V.V.~Flambaum, Proc. Phys. Scripta
46 (1993) 198.

\bibitem{spins}  B.~Georgeot and D.L.~Shepelyansky, Phys. Rev. Lett.
81 (1998) 5129.

\bibitem{FI97}  V.V.~Flambaum and F.M.~Izrailev,
Phys. Rev. E 56 (1997) 5144.

\bibitem{I00} F.M.~Izrailev,
in {\it Proceedings of the International  School of Physics
"Enrico Fermi"}, Course CXLIII, Varenna 20-30 July, eds.
G.~Casati, I.~Guarneri and U.~Smilansky, IOS Press, 2000,
pp.371-430.

\bibitem{kota} W.K.B.~Kota, Phys. Rep. 347 (2001) 223.

\bibitem{our} G.P.~Berman, F.~Borgonovi, F.M.~Izrailev, and V.I.
~Tsifrinovich,  Phys. Rev. E 64 (2001) 056226; Phys. Rev. E 65
(2001) 015204.

\bibitem{DISS05} M.I.~Dykman, F.M.~Izrailev, L.F.~Santos, and M.~Shapiro,
Phys. Rev. A 71 (2005) 012317; L.F.~Santos, M.I.~Dykman, M.
~Shapiro, and F.M.~Izrailev, Quantum Information and Computation,
5 (2005) 335.

\bibitem{FIC96}  V.V.~Flambaum, F.M.~Izrailev, and G.~Casati,
Phys. Rev. E, 54 (1996) 2136; V.V.~Flambaum and F.M.~Izrailev,
Phys. Rev. E 55 (1997) R13.

\bibitem{I01} F.M.~Izrailev, Physica Scripta T90 (2001) 95.

\bibitem{FGI96} V.V.~Flambaum, G.F.~Gribakin and F.M.~Izrailev, Phys. Rev. 53
(1996) 5729.

\bibitem{FI99}  V.V.~Flambaum and F.M.~Izrailev, cond-mat/9812417.

\bibitem{BM74} A.~Bohr and B.~Mottelson, in: {\it Nuclear
Structure}, Benjamin, NY, 1974.

\bibitem{FI01b} V.V.~Flambaum and F.M.~Izrailev, Phys. Rev. E
64 (2001) 036220.

\bibitem{FI01a} V.V.~Flambaum and F.M.~Izrailev,  Phys. Rev. E 64 (2001) 026124.

\bibitem{FI00} V.V.~Flambaum and F.M.~Izrailev, Phys. Rev. E 61 (2000) 2539.

\bibitem{W55}  E.P.~Wigner,
Ann. Math. 62 (1955) 548; Ann. Math. 65 (1957) 203.

\bibitem{FCIC96}  Y.V.~Fyodorov, O.A.~Chubikalo, F.M.~Izrailev and
G.~Casati, Phys. Rev. Lett. 76 (1996) 1603.

\bibitem{ber1} G.P.~Berman, G.D.~Doolen, G.D.~Holm, and V.I.~Tsifrinovich,
Phys. Lett. A 193 (1994) 444.

\bibitem{dima} B.~Georgeot and D.L.~Shepelyansky, Phys. Rev.
Lett. 81 (1998) 5129; Phys. Rev. E. 62 (2000) 3504; ibid, p.~6366.

\bibitem{BBIT01a} G.P.~Berman, F.~Borgonovi, F.M.~Izrailev, and
V.I.~Tsifrinovich,  Phys. Lett. A 291 (2001) 232.

\bibitem{BBCo02} G.P.~Berman, F.~Borgonovi, G.~Celardo, F.M.~Izrailev,
and D.I.~Kamenev, Phys. Rev. E 66 (2002) 056206.

\end{thebibliography}
\end{document}